\documentclass[pdflatex,sn-mathphys-num]{sn-jnl}

\usepackage{graphicx}%
\usepackage{multirow}%
\usepackage{amsmath,amssymb,amsfonts}%
\usepackage{amsthm}%
\usepackage{mathrsfs}%
\usepackage[title]{appendix}%
\usepackage[table]{xcolor}
\usepackage{textcomp}%
\usepackage{manyfoot}%
\usepackage{booktabs}%
\usepackage{algorithm}%
\usepackage{algorithmicx}%
\usepackage{algpseudocode}%
\usepackage{listings}%
\usepackage{ulem}

\begin{document} 


\title[Article Title]{Advances in Surrogate Modeling for Biological Agent-Based Simulations: Trends, Challenges, and Future Prospects}


\author*[1]{\fnm{Kerri-Ann} \sur{Norton}}\email{knorton@bard.edu}
\equalcont{These authors contributed equally to this work.}

\author[2,3,4,5]{\fnm{Daniel} \sur{Bergman}}\email{dbergman1@som.umaryland.edu}
\equalcont{These authors contributed equally to this work.}

\author[6]{\fnm{Harsh Vardhan} \sur{Jain}}\email{hjain@d.umn.edu}
\equalcont{These authors contributed equally to this work.}

\author[7]{\fnm{Trachette} \sur{Jackson}}\email{tjacks@umich.edu}
\equalcont{These authors contributed equally to this work.}

\affil*[1]{\orgdiv{Computational Sciences Program}, \orgname{Bard College}, \orgaddress{\street{30 Campus Road}, \city{Annandale-on-Hudson}, \postcode{12504}, \state{NY}, \country{USA}}}

\affil[2]{\orgdiv{Institute for Genome Sciences}, \orgname{University of Maryland School of Medicine}, \orgaddress{\street{655 W. Baltimore Street}, \city{Baltimore}, \postcode{21201}, \state{MD}, \country{USA}}}

\affil[3]{\orgdiv{Greenebaum Comprehensive Cancer Center}, \orgname{University of Maryland School of Medicine}, \orgaddress{\street{22 S. Greene Street}, \city{Baltimore}, \postcode{21201}, \state{MD}, \country{USA}}}

\affil[4]{\orgdiv{Institute for Health Computing}, \orgname{University of Maryland School of Medicine}, \orgaddress{\street{6116 Executive Boulevard}, \city{North Bethesda}, \postcode{20852}, \state{MD}, \country{USA}}}

\affil[5]{\orgdiv{Department of Pharmacology and Physiology}, \orgname{University of Maryland School of Medicine}, \orgaddress{\street{655 W. Baltimore Street}, \city{Baltimore}, \postcode{21201}, \state{MD}, \country{USA}}}

\affil[6]{\orgdiv{Department of Mathematics \& Statistics}, \orgname{University of Minnesota Duluth}, \orgaddress{\street{1117 University Dr}, \city{Duluth}, \postcode{55812}, \state{MN}, \country{USA}}}

\affil[7]{\orgdiv{Department of Mathematics}, \orgname{University of Michigan}, \orgaddress{\street{530 Church Street}, \city{Ann Arbor}, \postcode{48109}, \state{MI}, \country{USA}}}


\abstract{Agent-based modeling (ABM) is a powerful computational approach for studying complex biological and biomedical systems, yet its widespread use remains limited by significant computational demands. As models become increasingly sophisticated, the number of parameters and interactions rises rapidly, exacerbating the so-called ``curse of dimensionality'' and making comprehensive parameter exploration and uncertainty analyses computationally prohibitive. Surrogate modeling provides a promising solution by approximating ABM behavior through computationally efficient alternatives, greatly reducing the runtime needed for parameter estimation, sensitivity analysis, and uncertainty quantification. In this review, we examine traditional approaches for performing these tasks directly within ABMs—providing a baseline for comparison—and then synthesize recent developments in surrogate-assisted methodologies for biological and biomedical applications. We cover statistical, mechanistic, and machine-learning-based approaches, emphasizing emerging hybrid strategies that integrate mechanistic insights with machine learning to balance interpretability and scalability.  Finally, we discuss current challenges and outline directions for future research, including the development of standardized benchmarks to enhance methodological rigor and facilitate the broad adoption of surrogate-assisted ABMs in biology and medicine.}





\maketitle

\section{INTRODUCTION}\label{sec1}

Agent-based models (ABMs) have become an essential computational tool for studying complex biological and medical systems, enabling the simulation of individual agents' interactions to capture emergent behaviors at the system level \cite{norton2016effects,shuaib2016multi,badham2018developing,rikard2019multiscale,west2023agent}.  However, despite their utility, ABMs typically suffer from high computational costs associated with simulating millions of agents, making parameter exploration, sensitivity analysis, uncertainty quantification, and model parameterization from real-world data extremely challenging \cite{ghaffarizadeh2018physicell,jain2022smore,bergman2024connecting}. 
As model complexity increases, so does the number of parameters, leading to the well-known ``curse of dimensionality'', rendering exhaustive search of the entire parameter space infeasible \cite{ghaffarizadeh2018physicell,jain2022smore,bergman2024connecting,norton2016effects,zhang2020validation,broniec2021guiding,perumal2020surrogate}.  Compounding these challenges, there are currently no universally accepted standards regarding how ABMs should be calibrated and analyzed~\cite{angione2022using}.

Surrogate modeling offers a promising solution to these computational burdens by producing computationally efficient representations of ABMs that closely approximate their behavior while substantially reducing runtime, enabling repeated runs in only a fraction of the original simulation time~\cite{blanning1975construction,regis2005constrained,o2006bayesian,asher2015review,jain2022smore,bergman2024connecting,angione2022using}.  Also referred to as metamodels or response surfaces, these surrogates enable rapid parameter sweeps, optimization, and uncertainty quantification without requiring exhaustive simulation runs~\cite{blanning1975construction,regis2005constrained,o2006bayesian,asher2015review}. Recent advances, including machine learning-based surrogate models and hybrid approaches integrating machine learning with mechanistic methods, have further enhanced the applicability of surrogate models in ABM-driven research \cite{garzon2022machine,angione2022using,Lagergren2020,Hartman2023}.

\begin{figure}[h!]
    \centering
    \includegraphics{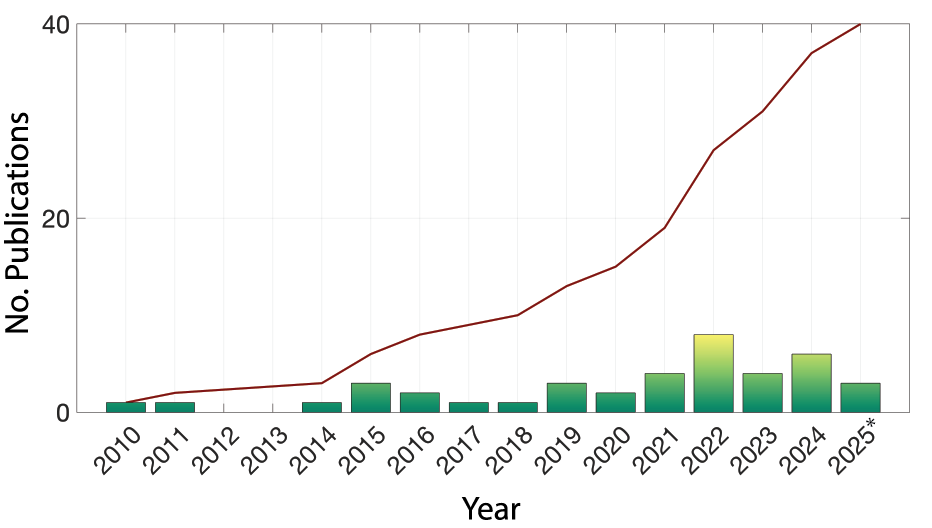}
    \caption{Publications identified through a PubMed search for titles or abstracts containing (‘surrogate model’ OR ‘SM’) AND (‘agent-based’ OR ‘ABM’ OR ‘individual-based’). The bar graph shows yearly results, while the line plot represents the cumulative number of publications over time.
    Asterisk indicates current year.}
    \label{fig:pubmedhits}
\end{figure}

This review explores the state-of-the-art surrogate modeling techniques for biologically- and medically-based ABMs, focusing on their applications in parameter estimation, identifiability, sensitivity analysis, and uncertainty quantification.
We begin by providing an overview of traditional statistical, machine-learned, and mechanistic surrogate models, examining emerging methodologies such as Biologically Informed Neural Networks (BINNs) and Universal Physics-Informed Neural Networks (UPINNs), and discussing the strengths and limitations of different approaches. 
Figure~\ref{fig:pubmedhits} highlights the rapidly growing interest in surrogate-based ABM simulations for biological or medical applications, showing both the annual number of published journal articles and the cumulative total of such publications over time, based on PubMed entries that mention ``agent-based models” and ``surrogate models” in the Title or abstract.  By synthesizing recent developments, we aim to guide researchers in selecting appropriate surrogate modeling techniques to enhance the scalability and interpretability of ABM-based investigations in biology and medicine.

We then discuss how surrogate models facilitate efficient parameter estimation in ABMs, enabling researchers to efficiently calibrate models against experimental or observational data with significantly reduced computational overhead.  These methods effectively address the challenges posed by high-dimensional parameter spaces and the associated computational intractability, thereby improving parameter identifiability.  Finally, we provide an overview of how surrogate models can enhance sensitivity analysis by quickly and accurately identifying key parameters that exert the strongest influence on ABM outputs.  We also examine their role in uncertainty quantification, which is critical for assessing model accuracy, robustness, and overall credibility.

We present a consolidated overview of surrogate-assisted ABM modeling with applications in biology, medicine, and complex systems while also identifying key research gaps and proposing new directions.  In doing so, we aim to influence the trajectory of future research in surrogate-driven computational modeling.  Ultimately, this work seeks to advance the field by providing a structured, accessible, and comprehensive resource for researchers and practitioners aiming to enhance the efficiency, accuracy, and interpretability of ABM-based simulations.

\section{BASICS OF SURROGATE MODELING}


Surrogate Modeling refers to creating simplified models that approximate the behavior of complex, computationally expensive, or difficult-to-analyze systems \cite{alizadeh2020managing,diaw2024efficient}.  These models are often constructed based on data collected from simulations of the original high-fidelity model or experimental data \cite{diaw2024efficient,roy2021feature}.  They are designed to predict output with minimal computational cost while maintaining an acceptable level of accuracy \cite{angione2022using}.  Surrogate models are used in various fields where direct simulation or analysis of the full system is impractical due to time, cost, or complexity \cite{diaw2024efficient}.

\subsection{Importance in Computational Sciences}

In computational biology, ecology, medicine, and other disciplines, surrogate models play a crucial role by reducing the computational burden of parameter estimation as well as sensitivity and uncertainty analyses~\cite{sakurada2024synthesis, gherman2023bridging}.  Complex models in these fields often require significant computational resources and simulation time.  Surrogate models offer a faster alternative, enabling researchers to investigate larger parameter spaces within reasonable time frames~\cite{jain2022smore,bergman2024connecting}.  For example, in computational biology, surrogate models can significantly alleviate the effort required to simulate protein folding and genetic networks~\cite{hie2022adaptive,gruver2021effective,asgharzadeh2020nanofe,costello2018machine}, where capturing every molecular interaction can be computationally prohibitive otherwise.  By approximating key biochemical pathways and cellular processes, surrogate models allow researchers to predict biological responses under various conditions without resorting to exhaustive simulations~\cite{costello2018machine}.

Surrogate approaches also facilitate parameter estimation in complex biological systems, where direct simulation may be prohibitively time-consuming~\cite{renardy2018parameter,gherman2023bridging,cai2021surrogate,jain2022smore,bergman2024connecting}.  For example, neural population models capture brain activity using large-scale systems of equations that describe the interaction between neurons and synapses \cite{d2013realistic,murray2018biophysical}.  Surrogate models are used to estimate parameters in these models by matching outputs (such as local field potentials or firing rates) to experimental data from electroencephalograms (EEGs) or magnetoencephalograms (MEGs) \cite{tripp2015surrogate}.  This approach speeds up the parameter search process, enabling efficient calibration of large-scale brain models.  Surrogate models also support rapid sensitivity analysis and uncertainty exploration, which are essential for understanding how parameter variations influence outcomes~\cite{renardy2018parameter,wang2022recent,garbo2024multi,bergman2024smore}.  For example, in~\cite{renardy2018parameter}, the authors reduced the computational burden of simulating yeast polarization by constructing a surrogate model and then quantifying how uncertainties in key parameters affect polarization results.  Section 4 provides a comprehensive survey of surrogate-driven calibration of ABM parameters in biomedical applications.

In many real-world applications, such as optimizing drug dosages in medicine or conservation strategies in ecology, surrogate models enable the rapid exploration of vast parameter spaces for the purposes of optimization that would otherwise be computationally prohibitive~\cite{oloulade2023cancer,willem2014active}.  For example, in \cite{willem2014active}, the authors introduced an iterative surrogate modeling process that systematically analyzes typical and extreme scenarios in an individual-based model of influenza transmission and a deterministic dynamic model for varicella-zoster virus (VZV) vaccination.  This process allowed for identifying key parameters that drive disease outcomes, helping to reduce dimensionality and decision uncertainty in model predictions.
The methodology demonstrated improved system understanding in both cases and provided insights for refining the model parameters to enhance policy decisions.  Section 5 provides a comprehensive survey of surrogate-driven sensitivity analysis of ABM parameters in biomedical applications.

\subsection{Types of Surrogate Models}

Several types of surrogate models are used across various domains in the biological sciences.  Below, we review statistical, mechanistic, and machine-learning approaches.  

\subsubsection{Statistical Surrogate Models}

Polynomial regression is one of the simplest forms of surrogate modeling, where the relationship between inputs and outputs is approximated by a polynomial function~\cite{zhao2022polynomial, Saman2012}.  It fits a curve or surface to data generated by a high-fidelity model~\cite{Saman2012,Li2021}, making it suitable when the variable relationships are sufficiently smooth and can be described by polynomial terms of varying degrees~\cite{Saman2012,Li2021,Kang2011,tu1999}.  Although direct applications of polynomial regression in biological ABMs are not extensively documented, this methodology has been used to approximate complex biological processes in quantitative systems oncology~\cite{meyer2023}.

Another popular approach, Kriging, is a powerful statistical method used in surrogate modeling and spatial analysis, particularly in geostatistics, engineering, and computer simulation applications~\cite{krige1951statistical, kleijnen2009kriging}.  It is named after South African mining engineer Danie Krige, and it is commonly used for interpolation and prediction of complex functions, often in cases where direct simulation or data collection is expensive or time-consuming~\cite{krige1951statistical, kleijnen2009kriging}.  Kriging is essentially a form of Gaussian process regression \cite{yang2019physics} used for predicting unknown values of a function at certain points based on known values at other points.  The core idea of Kriging is to model the underlying function as a realization of a stochastic process with a certain degree of smoothness and correlation structure.  It uses a weighted average of nearby observed data points to predict the value at an unobserved location, with the weights depending on the spatial structure of the data \cite{tuo2020kriging,Cressie1990}.

Kriging often achieves high accuracy, especially when the data exhibit strong spatial or structural correlation.  Its ability to estimate uncertainty alongside predictions is particularly valuable in surrogate modeling and design optimization \cite{Han2016}.  Limitations of the approach include the fact that as the number of data points increases, the Kriging process can become computationally expensive due to the need to invert large covariance matrices \cite{bouhlel2016improved}.  Kriging also requires good knowledge of the data's underlying covariance or correlation structure.  The results may be inaccurate if this is incorrectly specified \cite{cressie1987nonparametric}.  In the context of ABMs, Kriging has been used for surrogate modeling in several fields, including business, finance, and economics \cite{Barde2017, dosi2017}. However, we could not find instances of its application to biological or biomedical ABMs in the current literature.

Despite their utility, statistical surrogate models such as polynomial regression and Kriging face notable limitations when applied to complex biological ABMs.  High-dimensional parameter spaces often increase computational overhead, and small modeling errors can compound over time, particularly for highly nonlinear mmultiscaleprocesses.  Moreover, many statistical techniques assume smooth or stationary behaviors, which may not hold in biologically heterogeneous systems that feature discontinuities, tipping points, or emergent behaviors.  These constraints underscore the need for more advanced or hybrid approaches.

\subsubsection{Machine Learning Surrogate Models}\label{sec:machine_learning_sms}

Neural networks have become a preferred method for surrogate modeling, especially when dealing with highly complex nonlinear systems \cite{singh2024framework,garzon2022machine}.  A neural network surrogate model is a data-driven approach that approximates the behavior of a computationally expensive model and is based on input-output relationships derived from a set of training data \cite{garzon2022machine,angione2022using}.  These networks consist of layers of interconnected neurons that process input data and learn patterns by adjusting the weights of these connections \cite{kufel2023machine}.  In a surrogate modeling context, the network is trained to map inputs (e.g., physical parameters) to outputs (e.g., simulation results or experimental data)~\cite{zhou2020surrogate}.  Neural networks excel at capturing complex, nonlinear relationships, making them ideal for surrogate modeling of systems with large numbers of input variables with complex interdependencies~\cite{zhao2023surrogate,ren2022tutorial}.   Once trained, they can provide near-instant predictions for new input cases, rendering them particularly valuable for real-time applications and optimization processes~\cite{lepakshi2022machine,garzon2022machine}. 

For example, in~\cite{larie2021use}, a neural network surrogate was trained to predict the time evolution of an ABM modeling sepsis, a serious medical condition.  The surrogate model enabled rapid forecasting of disease progression--a critical factor for timely medical interventions.  Similarly, Pereira et al.~\cite{pereira2021deep} trained a deep neural network on data generated from ABM simulations to reduce the runtime of an infectious disease ABM.  The surrogate allowed for faster computations than multiple full-scale ABM runs and, crucially,  maintained consistent runtime regardless of the number of agents.  In another application, Cess and Finley~\cite{cess2020multi} developed a mmultiscaleABM of macrophages and T cell interactions within a tumor microenvironment to study the impact of macrophage differentiation on T cell-mediated tumor elimination.  They employed a neural network surrogate to approximate the detailed intracellular mechanistic model--specifically, the ordinary differential equations governing macrophage intracellular dynamics--thereby avoiding the substantial computational costs of evaluating these equations for every agent at every time step.  By integrating the surrogate into the ABM, they preserved biological detail, achieving over 98\% accuracy whilst significantly reducing computational complexity.

Nevertheless, neural networks also present challenges~\cite{bengio2009}.  Training a neural network can be computationally expensive, especially for deep architectures or large datasets~\cite{hinton2006}.  Without proper regularization, neural networks may overfit the training data, leading to poor generalization to new inputs~\cite{srivastava2014}.  Additionally, these models typically require substantial amounts of high-quality training data, which may not always be available in certain applications~\cite{halevy2009}.

Support Vector Machines (SVMs) are supervised machine learning algorithms used for both classification and regression tasks~\cite{cortes1995support}.  In classification, SVMs identify an optimal hyperplane that separates classes with the maximum possible margin~\cite{montesinos2022support,ardeshir2021support}, whereas in regression tasks, they aim to fit a line (or hyperplane in higher dimensions) so that most data points lie within a specified margin~\cite{montesinos2022support,ardeshir2021support}.  The goal is to minimize errors and achieve good generalization.  SVMs can serve as effective surrogate models because, once trained, they are relatively computationally inexpensive, can approximate highly nonlinear relationships, and are robust to overfitting--especially in high-dimensional spaces~\cite{ciccazzo2014support}.  Their reliance on only the data points closest to the decision boundary (support vectors) renders them less sensitive to outliers~\cite{kanamori2017breakdown}.  Findings in~\cite{angione2022using} suggest that SVMs can effectively approximate ABM outputs, thereby facilitating more robust sensitivity analyses and reducing computational costs.  Additionally, the framework proposed by~\cite{perumal2022surrogate} integrates different sampling methods and surrogate models, including SVMs, to parameterize an infectious disease ABM, with results indicating that SVM-based surrogates can achieve high accuracy and efficiency in approximating the model's behavior. 

However, several limitations affect the performance of SVMs as surrogate models~\cite{guido2024,sayeed2024,zhang2016}.  For example, when working with large datasets, the number of support vectors can grow significantly, making the model memory-intensive and slower to evaluate, and training times become prohibitive compared to other surrogate approaches like neural networks~\cite{shalev2007}.  Moreover, the choice of kernel and associated hyperparameters heavily influences SVM performance, and no single kernel is optimal for all problems.  Selecting an appropriate kernel often requires extensive domain knowledge and experimentation, and an unsuitable kernel can lead to poor approximations~\cite{mehmet2011}.  The inherent ``black-box" nature of SVMs, particularly when complex kernels are used, can limit interpretability--a drawback in applications where understanding underlying model relationships is critical, such as certain scientific or engineering fields~\cite{mehmet2011, guido2024}.


Decision trees are nonparametric supervised learning algorithms that create training models that can be used to predict the class or value of the target variable by learning simple decision rules inferred from training data.  This method recursively splits data based on feature values, resulting in a tree-like structure consisting of nodes (decision points), branches (decision outcomes), and leaves (final predictions)~\cite{strobl2009introduction}.  Their interpretability, capacity to capture nonlinear relationships, capacity to provide clear and understandable decision rules, and ability to highlight feature importance make them attractive as surrogate models. ~\cite{blockeel2023decision}.  

Ensemble methods, such as Random Forests and Gradient Boosting, build multiple decision trees and combine their outputs--using techniques like averaging for regression or majority voting for classification--to enhance predictive performance and mitigate overfitting \cite{jun2021comparison}.  When using the Random Forests method, for example, multiple subsets of data are randomly sampled with replacement, and a decision tree is trained on each subset.  Predictions are averaged (for regression) or determined by majority vote (for classification) \cite{jun2021comparison}.  Gradient Boosting, on the other hand, is another ensemble technique that builds trees sequentially, where each new tree corrects the errors of the previous ones \cite{jun2021comparison}.  These approaches have been successfully applied as surrogates in various domains, including ecology, epidemiology, economics, and social sciences, allowing researchers to replace computationally expensive simulations with rapid predictions based on input-output pairs generated by complex models \cite{angione2022using}.  For instance, \cite{robertson2024bayesian} introduced a Random Forest-based technique to accelerate the evaluation of a COVID-19 transmission ABM in Chicago.  The surrogate model effectively approximated the ABM's behavior, enabling faster scenario analyses for interventions like vaccination and lockdown policies. 

Despite these advantages, decision trees and ensemble models have notable limitations.  Decision trees can become overly complex and sensitive to small data variations, leading to different tree structures that impair generalizability~\cite{blockeel2023decision}.  Ensemble methods, while improving accuracy, are generally less interpretable and more resource-intensive than a single decision tree~\cite{jun2021comparison}.  We refer the reader to~\cite{sivakumar2022innovations} for an informative review of how ABMs have been integrated with machine learning approaches across diverse biological contexts, ranging from multicellular and tissue-scale dynamics to human population-level epidemiology.  The authors discuss suitable ML methods for specific ABM applications by considering both the scale of the biological system and the characteristics of the available data used to constrain various properties of the ABM.

\subsection{Mechanistic Surrogate Models}

Mechanistic models are mathematical or computational representations that describe the underlying processes and physical laws governing a system's behavior~ \cite{stalidzans2020, Levenstein2023}.  These models are based on physical, chemical, and/or biological principles, incorporating known mechanisms like biochemical reactions, material balances, or cellular signaling pathways~\cite{ma2024}.  They capture the causal relationships within a system, making them valuable for predicting system behavior under different conditions or interventions~\cite{stalidzans2020, Levenstein2023}.  Moreover, mechanistic models can serve as surrogates for high-fidelity simulations or real-world systems that are otherwise too complex to analyze directly~\cite{fonseca2025optimal, gherman2023, jain2022smore}. 

Mechanistic models are often considered superior surrogates than black-box models because of their interpretability, generalizability, and ability to incorporate known system dynamics~\cite{gherman2023, metzcar2024, shaghayegh2022}.  Their structure allows for clear connections between inputs, mechanisms, and outputs, which helps researchers understand how and why the model produces specific predictions~\cite{ gherman2023}.  In contrast,  black-box models (e.g., neural networks or other machine-learning techniques) tend to be opaque in their operations and difficult to interpret.  While they may achieve high predictive accuracy, their internal workings (such as weights and layers) do not directly correspond to real-world system components in an easily interpretable way~\cite{ hassija2024, Rudin2019}.

Because mechanistic models incorporate system-specific principles, they generally generalize well across a range of conditions within the modeled system~\cite{stalidzans2020, Levenstein2023, ma2024}.  For instance, altering input conditions--like substrate concentrations or temperature--in a biochemical pathway model does not require retraining; the model continues to produce accurate predictions within a new context because the underlying causal relationships remain valid.  On the other hand, black-box surrogate models are often limited to the specific data on which they were trained~\cite{JEONG2024, Rudin2019}.  Changes in operating conditions can lead to inaccuracies, as these models may struggle to extrapolate outside their training domains~\cite{sacks1989}.  Additionally, mechanistic models typically require less data for accurate calibration due to their reliance on established equations or system dynamics, whereas black-box models often demand large datasets to achieve similar accuracy, particularly when working with complex systems~\cite{JEONG2024, Rudin2019, sacks1989}.

Mechanistic models offer flexibility by being customizable to incorporate domain-specific knowledge, which can improve model fidelity and ensure alignment with real-world behaviors~\cite{stalidzans2020, Levenstein2023, ma2024}.  For instance, researchers can directly incorporate kinetic rate laws or binding affinities based on empirical studies when modeling drug interactions.  In contrast, integrating prior expert knowledge into black-box models usually requires extensive modification.  They operate primarily on patterns learned from data, so embedding domain-specific knowledge requires complex engineering and does not naturally align with the model’s structure~\cite{JEONG2024, Rudin2019, sacks1989}.

Despite their numerous advantages, especially in terms of interpretability and robustness, mechanistic surrogate models also face certain limitations compared to other surrogate modeling approaches~\cite{fonseca2025optimal}.  For example, building these models requires in-depth system knowledge--including the formulation of governing equations and accurate parameter values--which may not always be available or easily derived~\cite{fonseca2025optimal, baker2018}.  Calibrating mechanistic models to experimental or observational data can be especially challenging when data are sparse and noisy or when key parameters are unknown or difficult to measure directly~\cite{gherman2023}.  Furthermore, mechanistic models are often tailored to specific conditions, and increasing their complexity (e.g., more variables or interactions) can quickly make them unwieldy.  As a result, they may struggle to represent high-dimensional systems with many interacting components~\cite{fonseca2025optimal, baker2018}.  Nonetheless, mechanistic surrogates remain a promising approach, particularly when combined with hybrid or more scalable modeling strategies.  Section 4.3 provides successful examples of mechanistic surrogate-driven calibration of ABM parameters.

\section{EMERGING APPROACHES}\label{sec:emerging}
Recent advances in surrogate modeling have led to the development of hybrid neural network architectures that embed domain-specific knowledge directly into their structure~\cite{EGHBALIAN2023}. Two such promising methodologies discussed below are Biologically Informed Neural Networks (BINNs) \cite{Lagergren2020} and Universal Physics-Informed Neural Networks (UPINNs) \cite{RAISSI2019, podina2023}. By integrating the interpretability and generalizability of mechanistic models with the flexibility and scalability of machine learning, these approaches have the potential to address key limitations of existing surrogate methods.

\subsection{Biologically Informed Neural Networks (BINNs)}

Biologically Informed Neural Networks (BINNs) are neural network architectures that integrate prior biological knowledge—such as mechanistic interactions, conservation laws, and physiological constraints—directly into their design and training objectives.  By incorporating these constraints, BINNs reduce the effective complexity of the surrogate model and enhance both the interpretability and biological plausibility of its predictions~\cite{Lagergren2020, Hartman2023, nardini2024}.  This approach allows researchers to leverage established domain expertise to guide the learning process, ensuring that the resulting model adheres to known biological principles.

ABMs in biology and medicine often involve numerous parameters with nonlinear and interdependent effects~\cite{jain2022smore,bergman2024connecting}.  BINNs address these challenges by enforcing known biological constraints during training, which reduces the search space for parameter estimation and enables more focused and efficient exploration~\cite{Lagergren2020}.  For instance, when modeling tumor growth \cite{nardini2024, wang2024}, a BINN can incorporate established information about proliferation and apoptosis.  This embedded knowledge filters out non-physiological regions of parameter space that might otherwise be considered by a purely data-driven approach.  As a result, the structured nature of BINNs often leads to a more parsimonious model with fewer effective degrees of freedom, thereby improving both the robustness and efficiency of surrogate model calibration~\cite{Lagergren2020, Hartman2023}.  

BINNs have been applied to surrogate modeling of ABMs that simulate tumor cell proliferation and tissue regeneration~\cite{nardini2024}.  This work employed BINNs to construct predictive partial differential equation (PDE) models from stochastic ABM data that describe collective migration processes.  The BINN framework integrated two multi‐layer perceptrons: one learns to approximate the spatiotemporal total agent density, while the other estimates a density‐dependent diffusion rate.  Together, these components were trained to satisfy a diffusion PDE that governs the ABM behavior.  The approach was demonstrated in several case studies drawn from cell biology experiments, including barrier and scratch assays that mimic wound healing and tumor invasion.  By accurately forecasting ABM behavior at fixed parameter values and, when coupled with multivariate interpolation, predicting outputs at new parameter settings, the BINN-guided PDE model provides an interpretable and efficient surrogate for exploring complex biological systems.  Although this work focuses on PDE-based models, BINNs can also be adapted to learn ODE representations when appropriate for the system dynamics.

\subsection{Universal Physics-Informed Neural Networks (UPINNs)}

Universal Physics-Informed Neural Networks (UPINNs) are a machine learning technique that leverages neural networks to solve complex physical problems by incorporating known physical laws, even when only limited data is available \cite{RAISSI2019, podina2023}.  UPINNs enable the network to learn not only the solution to the problem but also to infer unknown hidden terms within the governing differential equations--effectively ``discovering" parts of the underlying physical or biological system that might not be fully understood initially, thus enhancing prediction accuracy and providing deeper insights into complex systems~\cite{podina2023}.

UPINNs extend the principles of physics-informed neural networks (PINNs)~\cite{RAISSI2019} by introducing a parameterized framework that is ``universal" in the sense that it is flexible enough to approximate unknown components of differential operators across a wide range of systems without requiring a predetermined form~\cite{podina2023}.  While PINNs traditionally integrate physical laws (e.g., conservation equations) into the loss function~\cite{RAISSI2019}, UPINNs further adapt these methods to capture hidden dynamics~\cite{podina2023}, making them particularly valuable for ABMs, where system dynamics can vary widely across the parameter space.

Even though these methods are quite similar in that they both embed domain knowledge into neural network surrogates, BINNs are tailored to incorporate specific biological constraints to refine partially known mechanistic models, whereas UPINNs provide a more general framework capable of discovering unknown dynamics by learning hidden components of differential operators without assuming a predetermined form.

\subsection{Other Approaches}

Other approaches for deriving surrogate models via supervised learning are also under active development.  These are primarily focused on two methodologies: sparse regression and theory-informed neural networks~\cite{Lagergren2020}.  In the sparse regression framework, numerical methods such as finite differences or polynomial splines are used to denoise data and approximate derivatives, forming a library of nonlinear candidate terms that are then selected via sparsity-promoting linear regression to construct a parsimonious model~\cite{brunton2016discovering}.  However, this method often requires large amounts of training data and is sensitive to noise, as it assumes that the unknown function can be represented as a linear combination of candidate terms~\cite{lagergren2020learning}.  In contrast, theory-informed neural networks, such as PINNs, leverage deep neural networks to approximate the solution of a governing dynamical system while embedding known physical laws into the loss function, thereby simultaneously learning both the solution and the governing parameters.  However, this approach relies on prior knowledge of the underlying mechanistic differential equation \cite{RAISSI2019, Lagergren2020}. 

Together, these supervised learning methods offer promising avenues for surrogate model derivation, though each approach presents trade-offs in data requirements, noise robustness, and underlying model assumptions.  In particular, BINNs and UPINNs are redefining how surrogate models are constructed for ABMs in biology and medicine.  These approaches can help ensure that surrogate predictions remain consistent with biological realities by embedding domain knowledge directly into the learning process.  Furthermore, their potential to efficiently navigate high-dimensional parameter spaces and perform global sensitivity analyses is poised to accelerate hypothesis testing and guide experimental designs.

\section{SURROGATE MODEL APPLICATIONS IN ABM PARAMETER ESTIMATION}

Parameter estimation, or calibration, is the process of finding particular parameter sets that produce model outcomes consistent with observations~\cite{kurchyna2023can}.  The computational complexity inherent in ABMs has profound implications for parameter estimation processes.  Long runtimes render exhaustive exploration of the parameter space infeasible, posing significant challenges for traditional parameter estimation techniques requiring numerous model evaluations~\cite{perumal2020surrogate}.  Moreover, the stochastic nature of ABMs necessitates repeated simulations for each parameter set, amplifying the computational burden and further complicating parameter estimation.  Methods for ABM parameter estimation can be broadly classified into three categories: (1) qualitative estimation methods, (2) quantitative estimation methods using direct approaches, and (3) quantitative estimation methods using surrogate modeling.  A brief overview of these approaches is provided below.

\subsection{Qualitative Estimation Methods}
Qualitative estimation refers to the process of identifying ranges of model parameters that yield behavioral patterns consistent with observed phenomena rather than optimizing numerical error metrics.  One commonly used approach for qualitative parameter estimation in ABMs is pattern-oriented analysis, also referred to as indirect parameterization.  This method involves conducting parameter sweeps to identify patterns of ABM behavior that match observed patterns in real-world systems~\cite{grimm2013individual}.  First, ranges of potential values for uncertain parameters are identified, and then parameter values are sampled from this space to generate a set of possible parameter combinations.  Next, the ABM outputs for these combinations are evaluated against a predefined set of observed patterns, which serve as filters to distinguish acceptable parameterizations from unacceptable ones.  For example, pattern-oriented analysis has been successfully applied to ABMs simulating immune system dynamics \cite{folcik2007basic}, as well as to ecological models \cite{wiegand2003using}, to replicate the observed patterns of behavior and population dynamics in natural systems.

A similar but more statistically robust approach to parameter estimation in ABMs is Monte Carlo filtering \cite{saltelli2004sensitivity}.  Like pattern-oriented analysis, this method rejects sets of model simulations that fail to meet predefined performance criteria, thereby providing an objective procedure for parameter selection.  System behavior is defined qualitatively by imposing constraints or thresholds on permissible ranges for model variables, along with a binary classification of model outputs as either acceptable or unacceptable.  The process begins by sampling parameters from specified statistical distributions.  A large number of Monte Carlo simulations are then performed, each associated with a unique set of input parameter values, and the outputs are classified based on the specified behavior criteria.  To evaluate parameter importance, a nonparametric test, such as the two-sided Kolmogorov-Smirnov test, is used to compare the distributions of acceptable and unacceptable outputs for each factor.  This test computes a significance level for rejecting the null hypothesis (i.e., that the factor does not sufficiently distinguish between acceptable and unacceptable behaviors).  For instance, this approach has been applied to calibrating a food web model \cite{rose1991parameter}.

Building on these approaches, another method for parameter estimation integrates Bayesian model checking with simulated annealing to generate parameter sets that satisfy expert-defined behavioral criteria \cite{hussain2015automated}.  This approach was applied to an ABM of acute inflammation, in which 28 unknown parameters were estimated to ensure the model reproduced four dynamic behavioral criteria, the timing and dosages of bacterial lipopolysaccharide administration that produced the desired clinical outcomes.  The authors suggest that future adaptations may allow the algorithm to learn behavioral specifications directly from time-series data, although this would still fall short of directly fitting the time-series data itself.  Moreover, Bayesian methods may not always be appropriate because they rely on prior knowledge about the probability distributions of the modeled components, which is seldom available \cite{broniec2021guiding}.

While these qualitative estimation methods provide valuable insights by focusing on reproducing observed behavioral patterns, they are not without limitations.  Their reliance on subjective criteria and fixed thresholds can introduce biases in parameter selection and may overlook subtle quantitative nuances of system dynamics.  Moreover, the computational demands associated with extensive parameter sweeps and Monte Carlo simulations can become prohibitive, especially in high-dimensional parameter spaces.  Ultimately, if the goal is to accurately reproduce numerical data, parameter estimation must rely on quantitative methods. 

\subsection{Quantitative Estimation Methods Using Direct Approaches}

Quantitative estimation methods aim to directly calibrate ABMs to numerical data, by identifying specific values of parameters that minimize error between model output and data.  For instance, in \cite{gallaher2017hybrid}, Gallaher et al. proposed a hybrid method for estimating ABM parameters in a model of glioblastoma growth, demonstrating superior performance compared to random sampling, genetic algorithms and simulated annealing, all of which exhibited suboptimal convergence in their application.  Their method leverages a high-performance computing cluster (HPCC) to iteratively refine the parameter space.  At each iteration, a hybrid approach, combining genetic algorithms with weighted random sampling, narrows the search space to identify parameter sets within an acceptable error range.  This approach successfully trained 16 parameters to fit multidimensional data, including bulk tumor size measurements, cell population sizes, and single-cell data.  However, it required 5000 parameter space samples and substantial computational resources on the HPCC to achieve these results.

Open-source software packages have also been used to estimate parameters in ABMs from real-world data.  For example, OPTUNA \cite{optuna_2019}, a hyperparameter optimization tool, employs the Tree-Parzen Estimation algorithm—a Bayesian optimization technique—to efficiently search for optimal parameters by constructing a probabilistic model of the relationship between parameter configurations and model performance \cite{watanabe2023tree}.  OPTUNA has been applied to parameter estimation in an ABM of COVID-19 spread in New York and the UK \cite{krivorotko2022agent}.  However, this method also requires a potentially large number of ABM simulations.

Others have used particle swarm optimization (PSO) \cite{kennedy1995particle} to estimate parameters in ABMs, leveraging its ability to efficiently search high-dimensional parameter spaces by iteratively refining candidate solutions based on swarm intelligence.  For example, Tong et al. \cite{tong2015development} applied PSO to estimate key parameters in an ABM of the immune response to influenza A virus infection.  Their method first uses local regression (LOESS) to create a statistical mapping from ABM inputs to outputs.  Then, they apply PSO to optimize four parameters by fitting simulated outputs to experimental data.

Although this approach demonstrated reliability and efficiency, it faces challenges in high-dimensional input spaces because LOESS, like many machine learning algorithms, suffers from the curse of dimensionality, which can lead to significant variability in parameter estimates~\cite{taylor2013challenging}.  Additionally, PSO requires extensive sampling to identify optimal parameters, particularly when acceptable values lie in a small region of the input space.  In Tong et al.’s influenza A virus case study, 3,465 ABM runs (sampling 385 inputs and running each input nine times) were required to estimate just four parameters, highlighting the substantial computational cost.  In subsequent work, Li et al. \cite{li2017developing} integrated generalized additive models (GAMs) and history matching to reduce the parameter search space by approximately one-third, lowering the number of ABM runs to 1,200 while still estimating four parameters.  Nevertheless, multiple waves of history matching may be needed to further refine parameter estimates, again increasing computational demands.  The authors suggest that parallel computing (e.g., GPU acceleration) may alleviate these computational costs and render ABM-based parameter estimation more scalable.

The effectiveness and reliability of PSO were further tested by Vlad et al. \cite{vlad2024parameter} as part of a comprehensive study comparing multiple metaheuristic methods for parameter estimation in ABMs.  The authors evaluated Markov chain Monte Carlo (MCMC), particle swarm optimization (PSO), genetic algorithms (GA), surrogate modeling (SM), and chaos game optimization (CGO) on two ABMs: a simple four-parameter toy ABM of infectious disease spread and a more complex 26-parameter ABM of acute respiratory viral infections in a city of 10 million people.  Their findings identified MCMC as the most promising approach, particularly when an initial parameter set informed by domain knowledge is provided.  Specifically, they sampled 1,000 initial parameter sets, then performed three runs of up to 200 iterations each to account for the stochastic nature of both the ABM and the optimization algorithms.  In contrast, PSO, GA, and CGO generally performed well for the simpler ABM but encountered local minima and boundary issues in the higher-dimensional parameter space of the more complex ABM.  For instance, PSO required a large number of runs and adjustments to keep particles within valid ranges, while CGO reached its best value quickly but then stagnated at a local minimum.  

Other authors have also applied similar approaches to parameter inference in ABMs.  For example,~\cite{ajelli2016spatiotemporal} used Metropolis–Hastings MCMC to calibrate a three-parameter Ebola model against recorded deaths in Liberia, while~\cite{kattwinkel2017bayesian} developed a particle MCMC algorithm for Bayesian inference of nine parameters in a two-state ecological ABM.  However, as noted in~\cite{kattwinkel2017bayesian}, long runtimes and high storage demands can limit the broader applicability of MCMC-based methods.  Similarly, \cite{robin2024system} employed an Iterated Ensemble Adjustment Kalman Filter to infer two parameters in a network-based ABM of MRSA, finding that although the parameter space was narrowed, a unique solution remained elusive.  In \cite{calvez2005automatic}, genetic algorithms were proposed for exploring 2-dimensional parameter space in agent-based simulations of ant foraging.

Despite their methodological differences, the above approaches share some common challenges: they typically rely on a large number of ABM simulations, rarely quantify uncertainty in parameter estimates (with the exception of Bayesian methods), and can be highly sensitive to the complexity of the underlying ABM.  Moreover, one comparative study~\cite{korkmaz2019adaptive} found that although all of the metaheuristic algorithms they examined achieved acceptable accuracy across multiple ABMs, no single method consistently outperformed the others across all problems.  This underscores the ongoing need for more flexible and broadly applicable parameter estimation methods that minimize computational expense, offer robust uncertainty quantification, and accommodate diverse ABM structures and research goals.

\subsection{Quantitative Estimation Methods Using Surrogate Modeling}

By reducing the complexity of the original model — whether through statistical methods, reduced-order modeling, or other techniques — surrogate models enable more tractable parameter exploration and calibration.  Although surrogate modeling remains relatively underexplored in ABM parameterization, interest in such approaches is now growing due to their potential to significantly reduce the computational costs of parameter space exploration.

\subsubsection{Black-box Surrogate Models}

Surrogate modeling approaches typically rely on machine learning (ML) or deep learning.  For instance, in \cite{perumal2020surrogate}, Perumal and van Zyl compared three ML algorithms—XGBoost, decision trees, and support vector machines—for estimating seven parameters in a simple SIR-type ABM of virus spread.  XGBoost is an ensemble gradient-boosting algorithm that constructs multiple decision trees to improve predictive performance.
They demonstrated that these surrogate-assisted methods significantly outperformed random and quasi-random Sobol sampling, with XGBoost and decision trees providing the most accurate approximations of the real data.  In subsequent work \cite{perumal2022surrogate}, Perumal and van Zyl developed a more flexible ABM parameterization framework, which integrates an intelligent sampling method with ML surrogate modeling.  However, they noted that despite significant improvements, surrogate-assisted strategies could still encounter suboptimal convergence, requiring multiple iterations to ensure accurate solutions.

In \cite{jorgensen2022efficient}, the authors employed machine learning to perform Bayesian inference in two ABMs: a four-parameter brain cancer model and a five-parameter SIRDS model for infectious disease spread.  They begin by building a surrogate model that approximates the ABM output and then use standard sampling techniques on this emulator to infer posterior distributions on model parameters.  Three emulator techniques — deep neural networks (NN), mixture density networks (MDN), and Gaussian processes (GP) — are compared based on how accurately they reproduce the ABM’s output distribution across different training set sizes and statistical metrics.  For the cancer ABM, the neural network most accurately recovered the mean, median, and distribution width.  However, in the epidemiological SIRDS model, the mixture density network proved superior given a sufficiently large training set, while GPs generally underperformed in both cases.

The authors found that, while a direct simulation-based approach would need over 40 million runs of the ABM for either model, roughly 10,000 ABM runs sufficed to train the ML algorithms in both cases, reducing computation by several orders of magnitude.  Despite these successes, the authors emphasized that no single ML approach universally excels.  Instead, the best method depends on the nature of the problem and multiple techniques should be tested to find the optimal balance between computational cost and inference accuracy.

Notwithstanding the notable advantages of using ML surrogates for calibrating ABMs, several critical limitations remain.  First, the data requirements are substantial: typically, tens of thousands of ABM runs must be generated to train an accurate and robust ML model, making these approaches themselves computationally expensive for large-scale or highly detailed ABMs. Second, ML surrogates generally operate as black-box models; they optimize predictive performance but do not clarify how input variables interact to produce emergent outcomes.  This opacity can be problematic for modelers who want to understand the causal pathways that drive system-wide behavior — especially in biological systems, where mechanistic insight is often important.  In addition, ML surrogates may not generalize well beyond the parameter regimes and scenarios covered by the training data, leaving open the possibility of poor extrapolation if the ABM is run under new conditions.

\subsubsection{Explicitly Formulated Surrogate Models}

To overcome some of the above limitations, we have developed a new method for calibrating parameters in ABMs: Surrogate Modeling for Reconstructing Parameter Surfaces (SMoRe ParS)~\cite{jain2022smore,bergman2024connecting}.  Unlike traditional black-box approaches, SMoRe ParS constructs an explicitly defined surrogate model whose parameters are directly mapped to both the ABM’s input variables and experimental or clinical data.  A key guideline to constructing the surrogate is that it should preserve relevant mechanistic insights from the ABM, allowing it to be informed by both the ABM’s internal dynamics and observed phenomena.  This approach substantially reduces the number of required ABM runs to generate training data — often a central bottleneck in high-dimensional simulations.  By formulating explicit equations, SMoRe ParS  facilitates the interpretation of how parameter uncertainties in the ABM propagate to uncertainties in the data fit, thereby providing a distributional view of parameter estimates.

SMoRe ParS has been successfully applied to estimating parameters in ABMs of both 2D in vitro tumor growth assays \cite{bergman2024connecting} and 3D vascular tumor growth \cite{jain2022smore}, effectively handling (on the order of) ten parameters while integrating multidimensional real-world data.  For the 2D assay, twelve parameters were inferred using approximately 2,400 ABM simulations, whereas in the 3D model only two parameters were estimated, with simulations conducted at nine distinct input parameter values.  Although SMoRe ParS still faces challenges when applied to models with extremely high-dimensional parameter spaces and the identification or formulation of appropriate surrogate equations can be time-consuming, emerging deep learning methods (see section \ref{sec:emerging}) offer promising avenues to overcome these limitations.  Furthermore, by optimizing sampling strategies (e.g., using Latin hypercube sampling), the number of required training simulations can be made to scale linearly with the number of ABM parameters, thereby facilitating the exploration of larger and more complex parameter spaces.  Overall, by explicitly linking ABM parameters to experimental observations, SMoRe ParS provides a transparent and potentially more efficient calibration framework compared to traditional black-box alternatives.  Explicitly formulated surrogates also offer straightforward interpretability and can, in and of themselves, be informative. However, a significant limitation of this approach is the substantial time and effort required to derive mechanistic models, as well as the challenges involved in scaling them to complex biological systems.

\begin{figure}[h!]
    \centering
    \includegraphics{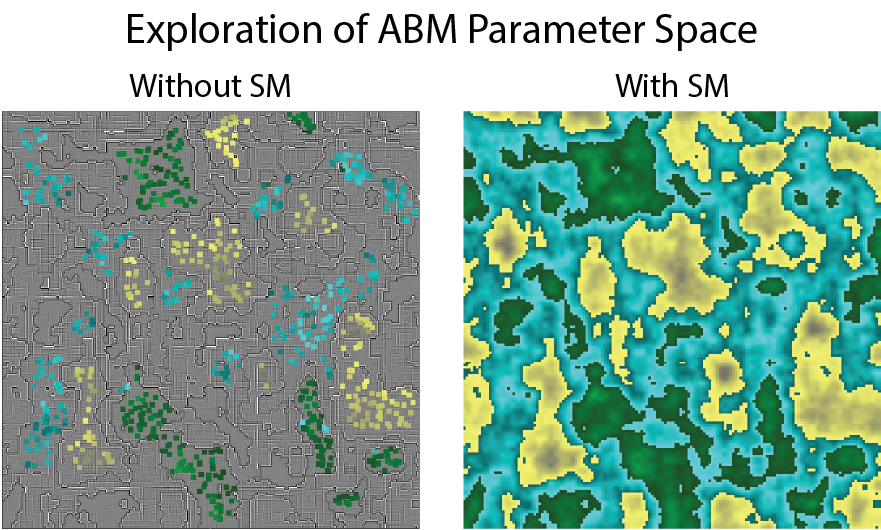}
    \caption{Surrogate models (SMs) enable more efficient exploration of the parameter space of an agent-based model (ABM).  While it is possible to accomplish only using the ABM, parameter exploration and uncertainty analysis can be accomplished much faster using an SM. The image shown was created using Perlin noise tiling, generated with a publicly available program~\cite{ThudlyReddit2019}.}
    \label{fig:exploremap}
\end{figure}

Figure~\ref{fig:exploremap} presents a schematic overview of the concepts discussed in Section 4. It illustrates the distinction between exploring the ABM parameter space directly—a detailed but slow approach that can result in sparse coverage—and using a surrogate model, which allows for broader strokes that quickly and efficiently yield a more comprehensive, albeit less granular, view of the parameter space.

\bigskip

\section{SURROGATE MODEL APPLICATIONS IN ABM UNCERTAINTY AND SENSITIVITY ANALYSIS}

\subsection{Uncertainty Analysis}

Uncertainty analysis involves identifying and classifying the various sources of uncertainty in a model and exploring how they affect the model’s behavior. The ultimate aim of such analysis is to provide insights into the reliability and robustness of model predictions~\cite{WALLACH2019209}.  Most uncertainty analysis techniques have traditionally been developed for deterministic models. In contrast, uncertainty analyses specifically addressing ABMs have remained relatively limited~\cite{marino2008methodology}; a PubMed search for titles or abstracts containing (“uncertainty analysis” OR “uncertainty quantification”) AND (“agent-based” OR “agent-based” OR “ABM”) yielded very few results prior to 2019, but returned 25 papers published since then, reflecting a growing interest in this field. 

In predictive mathematical modeling, uncertainty is typically categorized into two types—aleatory and epistemic—each addressing different sources of variability and error in the modeled system~\cite{Gansch2020SystemTV}. Aleatory uncertainty arises from inherent randomness and variability within the phenomena themselves. In ABMs, this randomness is often embedded in the stochastic rules governing agent decisions or interactions, causing outputs to vary even under identical initial conditions and parameter settings~\cite{kemkar2024towards,mcculloch2022calibrating}. Thus, aleatory uncertainty represents intrinsic variability that cannot be reduced by improved knowledge or additional data collection, and it is characterized through probability distributions of model outcomes. Epistemic uncertainty, in contrast, arises from gaps or limitations in our knowledge about the modeled system. In modeling terms, it represents uncertainty about appropriate parameter values or underlying model assumptions—commonly referred to as the ``unknown unknowns''—that can be reduced through improved data collection or methodological refinements~\cite{urbina2011quantification}. A model might struggle to capture the true dynamics of agents if it is based on incomplete data or relies on parameters estimated from sparse or imprecise measurements.

Recognizing these layers of uncertainty is crucial for accurately assessing the performance and predictive capability of mathematical models, as they collectively underscore both the intrinsic randomness of natural processes and the limitations of our current knowledge. However, in the context of ABMs, this process becomes especially challenging because, as noted in~\cite{marino2008methodology}, epistemic and aleatory uncertainties often become entangled when performing uncertainty analysis. As a result, distinguishing whether variability in model outputs stems from uncertain input parameters or from the model’s inherent stochasticity can prove difficult. 

In~\cite{marino2008methodology}, the authors propose a strategy to untangle aleatory uncertainty from epistemic uncertainty in ABMs, using a 12-parameter model of granuloma formation during Mycobacterium tuberculosis infection as a case study. First, they sampled parameter combinations from their underlying distributions. Next, they ran multiple simulations for each parameter combination and then averaged the simulation outputs. This replication-and-averaging approach aims to minimize the confounding effects of stochastic variability (i.e., aleatory uncertainty) on model output. The authors found that, when conducting a global sensitivity analysis (see section 5.3), this scheme effectively removed aleatory noise from the sensitivity calculations. However, they note that using a simple average to represent the distribution of outcomes is only reliable if the outputs are unimodally distributed around a central tendency.

This approach of running multiple ABM simulations per parameter set makes conducting uncertainty analysis computationally intensive. Surrogate models trained on averaged ABM behavior can potentially expedite these analyses by reducing the need for repeated full-scale simulations. Of course, computational costs will only be saved if the number of training simulations needed to develop the surrogate remains substantially lower than that required for direct analysis.

\subsection{Uncertainty Quantification}

An important component of uncertainty analysis is uncertainty quantification (UQ), which involves characterizing uncertainties associated with a model's inputs—including parameters, structural assumptions, and input data—and systematically propagating these uncertainties to assess their impact on the accuracy and reliability of the resulting outputs~\cite{smith2024uncertainty}.  However, the inherent stochasticity and computational complexity of ABMs make UQ particularly challenging for this class of models~\cite{baustert2017uncertainty,craig2001bayesian}.

UQ in modeling can be broadly divided into two categories: forward uncertainty quantification and inverse uncertainty quantification. Forward UQ entails propagating uncertainties in model inputs to predict their impact on the model’s outputs, typically by defining probability distributions for the possible outcomes associated with various input parameter sets \cite{pagani2021enabling}. This has significant overlap with sensitivity analysis~\cite{mcculloch2022calibrating} and is discussed further in the next subsection. 

In contrast, inverse UQ involves using experimental data to estimate the uncertainty in a model given a defined range of input parameters \cite{kakhaia2023inverse, prieg2020frameworks}.  We have discussed examples of inverse UQ applied to ABMs in biology and medicine in Section 4, such as \cite{ajelli2016spatiotemporal}, where a 3-parameter Ebola ABM was calibrated directly using MCMC for Bayesian parameter inference. Other studies~\cite{kakhaia2023inverse,jain2022smore} have proposed surrogate models to accelerate the inverse UQ process and make it computationally feasible. For instance, \cite{kakhaia2023inverse} employed Gaussian process regression as a surrogate for their 3D ABM of microscale arterial tissue, using Bayesian inference to calibrate seven parameters each across three submodels. In this approach, one Bayesian calibration iteration required at least 4000 draws of the model stress response—amounting to about 96 hours of wall clock time per arterial layer when using the original ABM. By contrast, the Gaussian process surrogate, trained on 1000 draws of stress coefficients, produced stress predictions in about 0.2 ms, reducing the total runtime by a factor of four. Once the surrogate was established, performing inverse UQ became nearly instantaneous. Although this surrogate-based method demonstrated significant gains, further improvements may be achievable through more mechanistic surrogates that require fewer training data. One successful example of such an approach is described in \cite{bergman2024connecting} and has been discussed in Section 4.3.

Best practices recommend performing both types of analysis—starting with inverse UQ to constrain parameter uncertainties, followed by forward UQ to assess output variability. This dual approach is especially important because a model may produce outputs that closely match experimental data while still lacking credibility if its governing parameters are poorly constrained or uncertain \cite{corti2021multiscale}.

\subsection{Sensitivity Analysis}

Sensitivity analysis (SA) explores which input parameters—and consequently, which underlying biological, physical, or real-world processes—most critically determine the outputs of interest~\cite{razavi2021future,saltelli2002sensitivity,saltelli2004sensitivity}. By systematically varying parameters and observing their effects on the distribution of model outputs, SA elucidates key input-output relationships, quantifies how parameter uncertainty contributes to output variability, and guides subsequent model refinements~\cite{ottaviani2024modern}. As a result, comprehensive SA can significantly reduce epistemic uncertainties arising from incomplete knowledge about model inputs, such as parameter values or initial conditions. 

SA can be local, which seeks to quantify the influence of perturbing a single input parameter on model output, or global, which evaluates output sensitivity to multiple input parameters varied simultaneously across their entire ranges~\cite{Zhou2008}. In highly non-linear models such as ABMs, with high-dimensional parameter spaces, global sensitivity analysis is the more obvious choice as the single parameter will not really be informative. A number of methods have been developed for conducting SA in parametric models, including variance-based methods, moment-independent techniques, one-at-a-time methods, Monte Carlo methods, and methods using spectral analysis (for recent reviews, see~\cite{iooss2015review,borgonovo2022sensitivity}). These methods have different objectives and come with different computational costs. For instance, the variance-based extended Fourier Amplitude Sensitivity Test (eFAST) or Sobol indices are also useful for factor prioritization as well as factor fixing. Factor prioritization identifies the most influential parameters, where fixing these would most reduce output uncertainty, while factor fixing aims at reducing the number of uncertain inputs by fixing unimportant factors. However, these methods are computationally expensive, scaling as $\sim10^3\times$ the number of parameters. Other regression-based methods like Partial Rank Correlation Coefficient (PRCC) offer intermediate computational efficiency and are effective for factor mapping—identifying input combinations producing output values above or below specified thresholds~\cite{marino2008methodology,saltelli2008global}. The computational expense of simulating complex models remains a major challenge in applying these methods to ABMs, with long run times rendering any meaningful global SA of such models unfeasible~\cite{sheikholeslami2019global}. For this reason, sensitivity analysis of complex, computationally expensive ABMs is often omitted or only partially performed~\cite{borgonovo2022sensitivity}. 

We begin by highlighting notable examples of direct global sensitivity analysis (SA) implementations in ABMs. One of the earliest comprehensive studies providing detailed methodology is~\cite{marino2008methodology}, where PRCC and eFAST methods were used to assess sensitivities of 12 input parameters in an ABM of granuloma formation during Mycobacterium tuberculosis infection. This study revealed significant computational challenges with eFAST, as it required substantially more simulations—53,456 total model runs compared to just 300 with PRCC. Moreover, eFAST exhibited a methodological artifact by incorrectly assigning aleatory variance to total-order sensitivity indices when applied to stochastic models. Consequently, Marino et al.~\cite{marino2008methodology} concluded that PRCC combined with Latin Hypercube Sampling (LHS) offers a more computationally efficient and reliable approach for analyzing sensitivity at specific time points, delivering comparable insights at a significantly lower computational cost. Another example~\cite{corti2020fully} coupled computational fluid dynamics (CFD) with ABMs to simulate hemodynamic-driven events in atherosclerosis, using PRCC-based global SA with seven input parameters sampled at only ten values. However, two drawbacks of the PRCC method are that it assumes a monotonic relationship between input parameters and the output of interest and assumes minimal correlation among input parameters—assumptions that may not hold in especially highly nonlinear models. Although eFAST overcomes these limitations, it does so at a significant computational cost. 

In their 7-parameter model of dengue transmission, \cite{kang2020cybergis} proposed a new method for performing global sensitivity analysis in spatially resolved ABMs, termed cyberGIS. CyberGIS integrates geographic information systems (GIS) with advanced computational infrastructures, facilitating efficient handling and analysis of large-scale spatial data. In their approach, the authors first identified key spatiotemporal scales, partitioned model outputs at each of these scales, and subsequently measured their mean and variance using forward uncertainty quantification. Finally, they applied a Sobol scheme to quantify global sensitivity indices. However, their method was computationally intensive, requiring approximately 16,000 ABM simulations. Another application of Sobol index-based SA in ABMs was presented in~\cite{gugole2021uncertainty}, where global SA was conducted on 2–4 parameters in a 14-parameter COVID-19 transmission model, requiring upwards of 10,000 simulations.

Overall, direct implementations of global SA in ABMs are relatively rare in the biological and biomedical literature, likely due to the significant computational burden associated with performing large numbers of simulations, especially for models with extended runtimes. Consequently, there remains substantial potential for surrogate modeling methods to accelerate global SA.

More recently, there has been interest in employing surrogate modeling to facilitate global SA in ABMs. In one study~\cite{ten2021use}, the authors proposed a two-step surrogate modeling approach using support vector machines (SVM) and support vector regression (SVR) for three different ABMs, a predator-prey model, a common-pool resource model, and a fishery model. First, SVM classification was used to divide ABM simulation data into distinct behavioral regimes based on predefined criteria (e.g., extinction versus persistence). Sensitivity indices were then computed to identify parameters that drive transitions between these regimes. Second, SVR regression was employed within each regime to assess quantitative variations in model output, thus identifying influential parameters governing detailed model behavior. Compared to conventional global SA methods—which aggregate sensitivity measures across the entire parameter space without considering qualitative differences—this regime-specific surrogate modeling approach provides more nuanced insights into parameter influence. Additionally, surrogate modeling significantly reduced the computational costs associated with performing global SA directly on ABMs. 

In another study, \cite{angione2022using} evaluated several machine learning models as surrogates for an ABM of social care provision in the UK, which simulates the interaction between the supply and demand of informal care. The surrogates were assessed based on their performance in conducting SA on 10 input parameters. Neural networks achieved the highest overall accuracy, especially when trained on larger datasets ($\sim1000$ samples), but required substantially more computation time. In contrast, gradient-boosted trees and nonlinear SVMs delivered slightly less accurate predictions but were faster to train and evaluate. 

These examples highlight the growing potential of surrogate models to support efficient and informative SA in ABMs. However, much work remains to develop surrogate modeling frameworks that are robust, scalable, and broadly applicable across diverse domains and model structures.


\section{Discussion}

Agent-based modeling in biology and medicine stands at the threshold of its next major leap forward. Computational and methodological advances have yielded increasingly sophisticated ABMs that can simulate biological systems with unprecedented granularity and realism~\cite{vieira2022computational}. However, to truly unlock their potential, these models must be rigorously connected to real-world biological and clinical data. Our review of the existing literature highlights the inherent challenges faced by modelers in achieving this connection. As demonstrated in various sections of this review, direct parameter estimation and uncertainty quantification in ABMs can be extremely computationally demanding. The stochastic nature of ABMs further compounds this issue, as it demands repeated simulations at each parameter set to accurately characterize intrinsic (aleatoric) variability.

\begin{figure}[h!]
    \centering
    \includegraphics{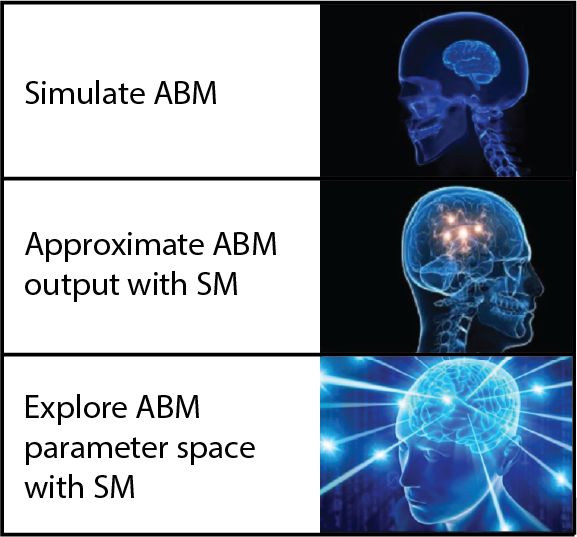}
    \caption{Means to explore agent-based models (ABMs) and analyze their output have grown rapidly. Surrogate models (SM) have emerged as a powerful tool for accomplishing these tasks. The open source `galaxy brain' meme images were taken from \cite{imageresizer_galaxy_brain}.}
    \label{fig:brainmeme}
\end{figure}

Surrogate modeling methods are emerging as a promising avenue to address these computational constraints, as illustrated in Figure~\ref{fig:brainmeme}. Surrogates can significantly speed up parameter calibration, sensitivity analysis, and uncertainty quantification by approximating ABM behavior at a fraction of the computational cost. The current literature highlights two primary surrogate modeling strategies: black-box machine-learning approaches and explicitly formulated, mechanistic surrogate models. Black-box surrogates offer substantial speed and scalability, but they often lack interpretability and require large volumes of training data. Recent studies suggest that the optimal machine-learning algorithm for surrogate modeling depends heavily on the specifics of the ABM being approximated, with neural networks generally demonstrating the strongest performance. However, these require extensive calibration datasets, partially offsetting their computational advantages. Mechanistic surrogates, meanwhile, offer greater transparency and generalizability, but their development can be labor-intensive, and they may not scale efficiently to highly complex or multidimensional ABMs. Our own experience~\cite{bergman2024connecting} also indicates that a single mechanistic surrogate often cannot accurately represent ABM outputs across the entire input parameter space, necessitating multiple surrogate formulations to effectively capture different parameter regimes.

Ultimately, the future of surrogate-augmented ABMs in biology and medicine likely involves integrating the strengths of both approaches into hybrid frameworks that leverage the flexibility of machine-learning methods alongside the interpretability and domain-specific insight provided by mechanistic modeling. Emerging techniques, such as biologically informed neural networks (BINNs) and universal physics-informed neural networks (UPINNs), exemplify this hybrid strategy by embedding known biological and physical constraints directly into data-driven surrogate architectures. Furthermore, promising directions such as equation learning~\cite{martina2021bayesian}—where mechanistically meaningful equations are inferred directly from data using sparse regression or deep learning—could bridge the gap between black-box and mechanistic surrogates, achieving a balance of interpretability, accuracy, and scalability.

To realize the full potential of surrogate modeling, the mathematical biology community needs standardized benchmarks to systematically evaluate and compare emerging methodologies. Currently, the absence of such benchmarks poses a notable limitation. Developing a consistent benchmarking framework would aid modelers in identifying and selecting surrogate modeling strategies best suited for their specific applications. To address these challenges, we suggest the following roadmap:

\begin{enumerate}

    \item \textbf{Curate a diverse set of ABMs.} A representative selection of ABMs should be identified spanning different model structures, parameter space dimensions, and simulated biological or medical processes. Each model should include a clearly defined dataset or a reproducible methodology for dataset generation to serve as training and evaluation data for surrogate models (SMs)
    
    \item \textbf{Define constraints on SM construction.} For methodologies that rely on user-defined SMs, standardized constraints should be established to ensure comparability. At a minimum, an upper bound on SM complexity, determined through approaches such as identifiability analyses~\cite{eisenberg2017confidence}, will be necessary to prevent overfitting and allow for meaningful comparisons across methods.
    
    \item \textbf{Establish resource constraints for training.} Benchmarks should incorporate a tiered system of computational thresholds, enabling assessments at various resource investment levels—from small-scale simulations on standard computing setups to high-performance computing (HPC) clusters with parallel processing and Message Passing Interface (MPI) support. This will help modelers assess methodology feasibility within their computational constraints, particularly in terms of the number of simulations required rather than raw wall-clock runtime alone.
    
    \item \textbf{Develop standardized benchmarking metrics.} Metrics should assess SM accuracy, successful  ABM parameter calibration, and the reliability of uncertainty quantification. Additionally, qualitative aspects such as ease of implementation, robustness, and adaptability to different ABMs should also be included. These metrics will help clarify the types of ABMs for which each methodology is most suitable, thereby guiding modelers in their methodological choices.

    \item \textbf{Establish a publicly accessible repository.} To ensure transparency and reproducibility, a public repository such as GitHub should be used to host standardized datasets, surrogate modeling scripts, and detailed documentation of benchmark tests. This resource will enable others to readily reproduce, evaluate, and adopt new methodologies, facilitating widespread community engagement and incremental improvements.
    
\end{enumerate}

To facilitate the creation of such benchmarks, we propose that researchers developing surrogate modeling techniques to augment their ABMs report their results in a standardized framework. Such standardized reporting will enable benchmarks to develop organically from community-driven efforts, providing a readily available set of test cases against which new methodologies can be evaluated. Specifically, we propose that each study includes the information outlined in Table~\ref{tab:benchmark_report} as far as possible. This standardization will ensure sufficient detail and transparency, enabling other researchers to objectively assess the relevance and performance of these surrogate approaches for their own modeling goals.

\renewcommand{\arraystretch}{1.2}
\begin{table}[]
    \centering
    \begin{tabular}{|l|p{4.5cm}|p{4cm}|}
        \hline
        \rowcolor[gray]{0.9}
        \textbf{Component} & \textbf{Description} & \textbf{Example} \\
        \hline
        ABM(s) & The ABM(s) on which the method is tested, including the number of input parameters & Well-established ABMs, e.g. PhysiCell hypoxia model and/or user-defined ABM \\
        \hline
        Data & The data used for training and/or validation & Experimental data and/or synthetic data \\
        \hline 
        Method(s) & The methods used to train the SM and/or ABM & PSO, MCMC, SMoRe ParS, Gaussian Process, etc. \\
        \hline
        Metric & Metrics used for training and validation as well as output of the method(s) & RMSE, credible intervals, sensitivity indices \\
        \hline
        Metametrics & Metrics to quantify ease of implementation & Number of simulations, CPU/GPU usage, HPC, MPI \\
        \hline
    \end{tabular}
    \caption{A set of features to report in any work using surrogate models (SMs) with agent-based models (ABMs) to build towards standardized benchmarks.}
    \label{tab:benchmark_report}
\end{table}

Developing standardized benchmarks for surrogate modeling in ABMs represents a transformative opportunity for the field. Clear and reproducible comparisons will enable modelers to confidently select suitable methodologies, reducing trial-and-error and enhancing the impact of their research. Benchmarks can foster cross-group learning, accelerating methodological progress and making new surrogate modeling approaches more accessible. By collectively defining and adhering to best practices, the field can ensure that innovations in surrogate modeling lead to broadly applicable, interpretable, and computationally efficient solutions. However, we recognize that establishing such benchmarks is inherently challenging, given the wide variety of ABM structures, differing computational resources among research groups, and the trade-offs intrinsic to surrogate modeling. A successful benchmarking effort will require iterative refinement as insights accumulate and methodologies evolve. Nonetheless, without these structured benchmarks, surrogate modeling applications to ABMs will remain largely ad hoc, limiting generalizability and slowing methodological advancement. By proactively addressing these challenges, the field can move toward a cohesive and scalable approach to fully unlock the potential of ABMs as predictive tools in biology and medicine.


\section*{Acknowledgments}

The authors acknowledge generous support from the Institute
for Computational and Experimental Research in Mathematics (ICERM) through
their Collaborate@ICERM program, the American Institute of Mathematics
(AIM) through their AIM SQuarREs program and the Mathematisches Forschungsinstitut Oberwolfach (MFO) through their Researchers in Pairs (RiP) program.  This work was supported by NIH/NCI U01CA243075 (TJ) and by NSF 2324818 (TJ, HVJ and K-AN).


\section*{Statements and Declarations}

\textbf{Competing interests:} The authors declare that they have no conflict of interest. \\

\noindent \textbf{Data Availability:} This manuscript has no associated data.

\bibliography{Review-article}
\end{document}